\documentclass[a4paper,preprintnumbers,floatfix,superscriptaddress,prl,twocolumn,showpacs]{revtex4-1}

\usepackage{amsmath, amsthm, amssymb}
\usepackage{graphicx}% Include figure files
\usepackage{dcolumn}% Align table columns on decimal point
\usepackage{bm}% bold math
\usepackage{hyperref} %hyperref package
\usepackage{braket}
\usepackage{color}

\newcommand{\vack}{\ket{vac}}

\newcommand{\figref}[1]{Fig.~\ref{#1}}

\DeclareGraphicsExtensions{.pdf,.png}

\begin{document}

%\title{Architechture for device-independent quantum key distribution with dipole-cavity memories}
\title{Device-independent quantum key distribution with spin-coupled cavities}

\author{Alejandro M\'{a}ttar}\altaffiliation[Corresponding author.\,\,\,]{alejandro.mattar@icfo.es}\affiliation{ICFO-Institut de Ci\`encies Fot\`oniques, Mediterranean Technology Park, 08860 Castelldefels (Barcelona), Spain}
\author{Jonatan Bohr Brask}\affiliation{ICFO-Institut de Ci\`encies Fot\`oniques, Mediterranean Technology Park, 08860 Castelldefels (Barcelona), Spain}
\author{Antonio Ac{\'i}n}\affiliation{ICFO-Institut de Ci\`encies Fot\`oniques, Mediterranean Technology Park, 08860 Castelldefels (Barcelona), Spain}\affiliation{ICREA-Instituci\'o Catalana de Recerca i Estudis Avan\c cats, Lluis Companys 23, 08010 Barcelona, Spain}

\date{\today}

\begin{abstract}
Device-independent quantum key distribution (DIQKD) guarantees the security of a shared key without any assumptions on the apparatus used, provided that the observed data violate a Bell inequality. Such violation is challenging experimentally due to channel losses and photo-detection inefficiencies. Here we describe a realistic DIQKD protocol based on interaction between light and spins stored in cavities, which allows a heralded mapping of polarisation entanglement of light onto the spin. The spin state can subsequently be measured with near unit efficiency. Heralding alleviates the effect of channel loss, and as the protocol allows for local heralding, the spin decay is not affected by the communication time between the parties, making Bell inequality violation over an arbitrary distance possible. We compute the achievable key rates of the protocol, based on recent estimates of experimentally accessible parameter values and compare to the other known DIQKD protocol, which is entirely optical. We find significant improvements in terms of key bits per source use. For example we gain about five orders of magnitude over a distance of 75km, for realistic parameter values.
\end{abstract}

\maketitle

%Introduction -- %
A fundamental task in cryptography is the distribution of a secret, shared key between two parties, Alice and Bob, which then enables them to communicate securely. Secrecy of the key means that the information of any eavesdropper about the key is negligible, which needs to be shown rigorously and ideally under as few assumptions as possible. Device Independent Quantum Key Distribution (DIQKD) \cite{acin2007,Pironio2009,barrett2005,mayers1998,masanes2011,pironio2012,Vazirani2012}, provides a setting in which such secrecy can be guaranteed without any assumptions whatsoever about the inner workings of the devices used in the protocol. The only prerequisite is that the observed data violate a Bell inequality \cite{bell1964}, however it must do so unconditionally, that is, without postselection on the observed outcomes. All tests of Bell inequalities so far for distances exceeding a few tens of meters suffer from the so-called detection loophole \cite{Pearle1970} -- the Bell inequality is not violated by the observed dataset as a whole, due to losses of the photons used in the experiments as well as imperfections in their detection. Usually, it is then assumed that the experimental runs with successful detection represent a fair sample of the entire dataset, the data analysis is restricted to this subset, and Bell inequality violation is restored. However, in the DIQKD setting where the channel and detectors may be under the control of a malicious eavesdropper, such an assumption is certainly not valid. It is therefore desirable to combat channel and detector losses in a different manner, which does not require postselection.

In an all-optical scheme, channel loss can be eliminated by means of noiseless linear amplification \cite{ralph2009,marek2010}, which allows the arrival of a transmitted photon to be heralded. A few years ago, a proposal for DIQKD based on this was put forward \cite{Gisin2010} and similar ideas have led to improvement of traditional QKD \cite{blandino2012} and proposals for loophole-free Bell tests \cite{brask2012}. Here we take a different approach, making use of the interaction between light and single spins in cavities. A Faraday-type spin-photon interaction has been shown to enable entanglement generation and swapping \cite{duan2004,Hu2008,hu2008_2,hu2009,hu2011} and has recently been proposed as the basis of a loophole-free Bell test \cite{Brunner2013}.

An important advantage of spin-cavity systems is that the spin state can be measured with near unit efficiency. At the same time, remote cavities can be entangled by mapping the entanglement from a pair of photons onto the spins in a heralded manner. The heralded arrival eliminates photon loss (a lost photon never arrives), and as the measurements settings for the spin measurement are only decided after successful heralding, the photon measurement constitutes a preselection that does not open any loopholes. Together, these facts lead to improvements of the attainable key rate and the possibility for Bell inequality violation over an arbitrary distance. In the following, we first describe the protocol in more detail, and then compute the key rates that can be achieved, taking into account realistic imperfections such as limited spin-photon coupling, spin decoherence, and optical detection efficiency.

\begin{figure}[h]
\begin{center}
\includegraphics[width = .95\linewidth]{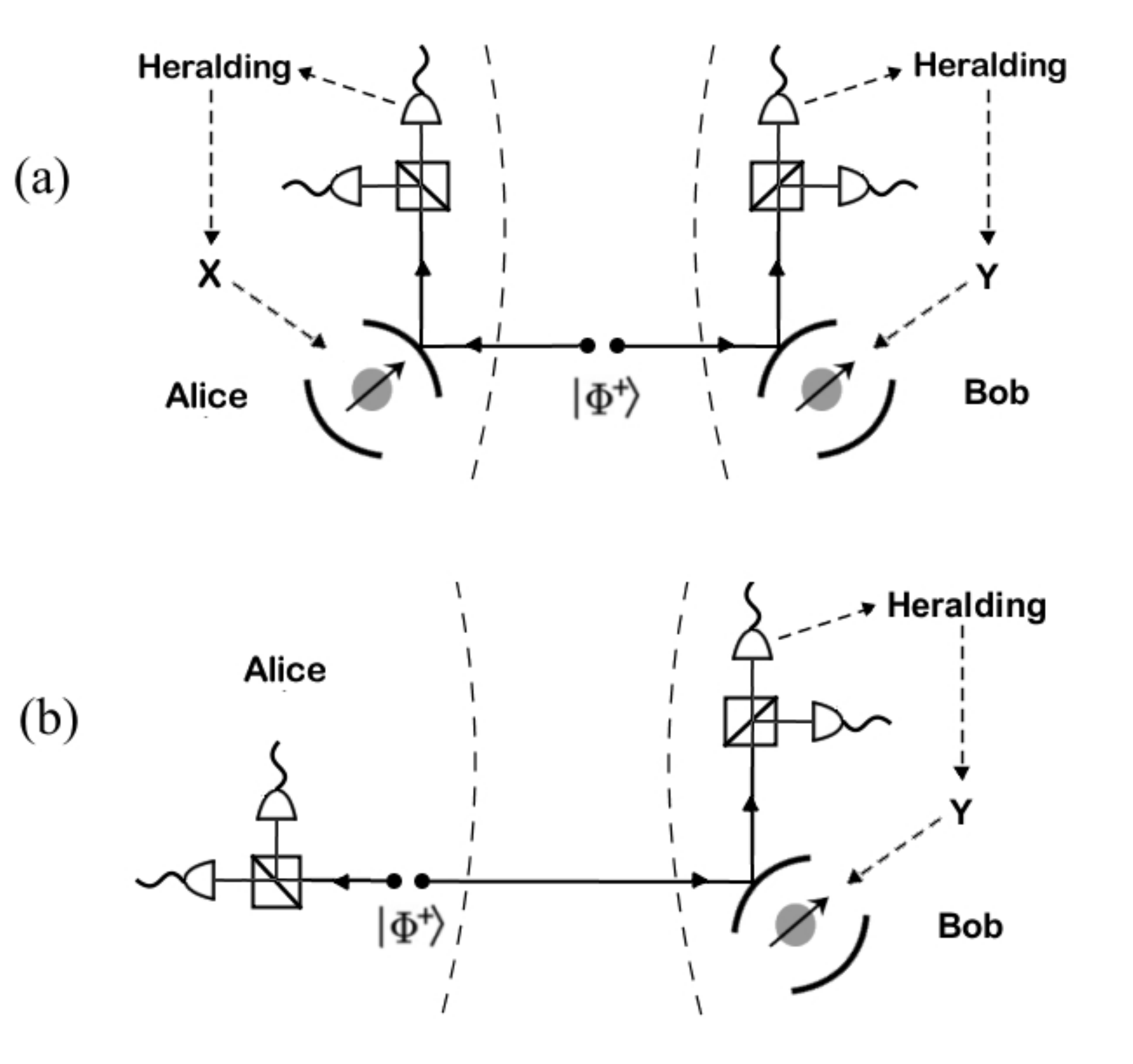}
\caption{\textbf{(a)} Symmetric protocol. A source emits entangled photon pairs. Each party holds a spin in a cavity, initialised in the state $(\ket{\uparrow}+\ket{\downarrow})/\sqrt{2}$. One photon interacts with each cavity and, if reflected, is measured in the $\ket{H}$, $\ket{V}$ polarisation basis. When such a heralding detection occurs on both sides, the photonic entanglement is mapped onto the spins. Once Alice observes the heralding photon, she chooses a basis and reads out her spin (similarly for Bob). The data from these spin measurements is used for the Bell test and key generation. \textbf{(b)} Asymmetric protocol. Bob proceeds as above, but Alice has no cavity. Instead, she directly measures the photon she receives in a polarisation basis of her choice. The data is generated from her outcome and the spin measurement of Bob. The source is located at Alice's side to minimise channel losses preceding her measurement.}\label{fig.setup}
\end{center}
\end{figure}

The architecture of our protocol is shown in \figref{fig.setup}. We will consider both symmetric and asymmetric variants of the protocol in which, respectively, both parties have cavities and a source of entangled photons is placed somewhere between them, or only Bob holds a cavity and the source is held by Alice who measures directly on her photon. The symmetric protocol closely resembles the Bell test setup of \cite{Brunner2013}. However, we note that while there heralding outcomes are communicated between the parties before the Bell test measurements are performed, no such communication is necessary with the strategies of the present schemes. That is, Alice does not need to learn Bob's heralding outcome before performing her measurement, and vice versa.

For the spin-photon interface, we consider the scheme proposed in \cite{Young2013} which works in the low Q-factor regime. It is a single-sided spin-cavity system characterized by four constants: $\kappa$, the outcoupling rate for intra-cavity light via the front mirror, $\kappa_s$, the decay rate of light into other loss modes (including absorption), $g$, the spin to cavity field coupling rate, and $\gamma$, the linewidth of the dipole transition. When $g^2 = \frac{\gamma (\kappa+\kappa_s)}{4}$ (resonance scattering \cite{Andreani1999}), any input photons resonant to the dipole-cavity system are scattered into loss modes, due to destructive interference between the input light and light scattered from the dipole. Thus, the presence of the spin strongly modifies the reflectivity of the cavity. The reflectivities for an empty cavity ($g=0$), and for a cavity resonantly coupled to the spin, for a field at zero detuning, are given by \cite{Hu2008}
\begin{equation}
\label{eq.refcoeffs}
r_c= \left| \frac{1-\kappa / \kappa_s}{1+\kappa / \kappa_s} \right| \hspace{1cm}  r_d=\frac{1}{1+\kappa / \kappa_s} .
\end{equation}
Similar expressions apply to other systems, such as atoms, and NV-centers \citep{Brunner2013}. Whether an incident photon will couple or not to the cavity spin (i.e.~see the cavity as empty or full) depends on the spin state as well as the photon polarisation. With a charged quantum dot in the cavity, the reflection coefficients for the joint circular polarisation and spin states are $r_d$ for $\ket{R,\uparrow}, \ket{L,\downarrow}$, and $r_c$ for $\ket{R,\downarrow}, \ket{L,\uparrow}$. Clearly, when the outcoupling $\kappa$ is small relative to the loss rate $\kappa_s$, all states transform the same, and there is no interaction between photons and spin. The ideal limit for our purposes is $\kappa \gg \kappa_s$ in which case $r_c \approx 1$ and $r_d \approx 0$.

To see how the spin-photon interaction is used in the protocol, consider the interaction of one (asymmetric protocol) or two (symmetric) spins initialised in the state $(\ket{\uparrow}+\ket{\downarrow})/\sqrt{2}$ with a Bell state of light $\ket{\Phi^+} = (\ket{RR} + \ket{LL})/\sqrt{2}$. We condition on reflection of the photon(s), and the unnormalised states then become
\begin{equation}
\ket{\psi^A} = \frac{1}{2}\left \{
\begin{array}{l}
\ket{H}\otimes[r_c\ket{\Psi^{+}}+r_d\ket{\Phi^{+}}]
\\+
\\ i\ket{V}\otimes[r_c\ket{\Psi^{-}}+r_d\ket{\Phi^{-}}]
\end{array}
\right. ,
\end{equation}
and
\begin{equation}
\ket{\psi^S}=\frac{1}{4}\left \{
\begin{array}{l}
\ket{HH}\otimes \left[(r_c^2+r_d^2)\ket{\Phi^{+}} + 2r_cr_d\ket{\Psi^{+}}\right]
\\+
\\i\ket{HV}\otimes (r_c^2-r_d^2)\ket{\Phi^{-}}
\\+
\\i\ket{VH}\otimes (r_c^2-r_d^2)\ket{\Phi^{-}}
\\-
\\\ket{VV}\otimes \left[(r_c^2+r_d^2)\ket{\Phi^{+}} + 2r_cr_d\ket{\Psi^{+}}\right].
\end{array}
\right. ,
\end{equation}
where we have separated the reflected photon(s) from the remaining state shared by Alice and Bob. $\ket{\Psi^\pm}$, $\ket{\Phi^\pm}$ denote the four Bell states. One sees that, for $\kappa \gg \kappa_s$, a measurement of the reflected photon(s) in the $\ket{H}$, $\ket{V}$ basis leaves the remaining state (of photon-spin or spin-spin for $\ket{\psi^A}$, $\ket{\psi^B}$ respectively) maximally entangled. For finite values of $\kappa/\kappa_s$ entanglement persists but is not maximal. The photonic measurement serves as a herald which primes the system for a subsequent Bell test. In the case of ideal detectors and no loss, the probabilities for each heralding outcome are $p_H = p_V = (r_c^2 + r_d^2)/4$ for the asymmetric and $p_{HH}=p_{VV}=\left[(r_c^2+r_d^2)^2+ 4r_c^2r_d^2\right]/16$, $p_{HV}=p_{VH}=(r_c^2-r_d^2)^2/16$ for the symmetric protocol. The probabilities for successful heralding are thus $p^A_{her} = p_{H} + p_{V} = (r_c^2 + r_d^2)^2/2$ and $p^S_{her} = p_{HH} + p_{HV} + p_{VH} + p_{VV} = (p^A_{her})^2$.

Note that the shared state after heralding depends on the measurement outcome. Thus a natural way for Alice and Bob to proceed is to communicate their heralding results to each other prior to the Bell test, and adapt the measurements of the test to the state they have. However, such communication requires a time $L/c$ where $L$ is the distance between the parties and $c$ is the speed of the signal. During this time the state will decohere, thus above some critical distance Bell inequality violation is no longer possible, and key distribution fails. To circumvent this problem, Alice and Bob can adapt the following strategy which does not require communication: whenever one of them observes a "V" herald, a $\pi$ phase-shift is applied to the spin. For the asymmetric protocol, the state after heralding is then always ($n_A$, $n_S$ are normalisation constants)
\begin{equation}
\label{eq.asymstate}
\ket{\varphi^A} = \left[ \ket{\Psi^{+}} + \frac{r_d}{r_c} \ket{\Phi^{+}} \right] / n_A ,
\end{equation}
For the symmetric protocol the states are
\begin{equation}
\label{eq.symstates}
\begin{alignedat}{2}
\ket{\varphi^S_0} & = \ket{\Phi^{+}} & & \hspace{0.5cm} \text{HV,HV} \\
\ket{\varphi^S_{+/-}} & = \left[ \ket{\Phi^{+}} \pm \frac{2r_cr_d}{r_c^2+r_d^2}\ket{\Psi^{+}} \right]/n_S & & \hspace{0.5cm} \text{HH/VV}
\end{alignedat}
\end{equation}
These states are not identical, however if $\kappa/\kappa_s$ is not too small they are close, and the Bell test can proceed by ignoring the heralding outcomes and considering their mixture, with $\ket{\varphi_{0,\pm}}$ weighted by $(p_{HV} + p_{VH})$, $p_{HH}$, and $p_{VV}$ respectively. Which strategy to adopt is just a matter of whether a higher key rate is extracted by communicating, thus sustaining more decoherence but performing the Bell test with optimal measurements, or by not communicating, sustaining less decoherence, but using suboptimal measurements. We find that for the parameter ranges relevant here, it is always advantageous to use the communication-free strategy.

The most important imperfection affecting the protocol is spin decoherence. To model it, we assume that noise on separate spins is independent (as they are far apart) and adopt a worst-case model of depolarising noise acting on a timescale $\tau$. That is, each spin is subject to a depolarising channel \cite{NielsenChuang}
\begin{equation}
\xi(\rho)=\sum_{i=0}^3\gamma_i\ \sigma_i\ \rho\ \sigma_i ,
\end{equation}
where $\sigma_0$ is the identity, $\sigma_i$ the Pauli matrices, $\gamma_0 = (1+3\exp(-t/\tau))/4$, and $\gamma_i = (1-\exp(-t/\tau))/4$ for $i=1,2,3$. Here $t$ is the time during which the spins decohere, which can be taken to be the time between heralding and the end of the spin measurement. For a communication-free strategy, this time is governed by the readout time $t_m$ (including possible electronic delays), while otherwise communication must be taken into account $t = t_m + L/c$ \footnote{Although the phase-shift operator $\sigma_z$ does not commute with the Kraus operators of the noise, the order of noise and phase shifts is nevertheless arbitrary, because $\sigma_z^2 = \sigma_0$ and $\sigma_z\sigma_i = -\sigma_i\sigma_z$ for $i\neq z$, which implies e.g.~for a phase shift on Alice's side $(\sigma_z\otimes\sigma_0)(\sigma_i\otimes\sigma_j)\rho(\sigma_i\otimes\sigma_j)(\sigma_z\otimes\sigma_0) = (\sigma_i\otimes\sigma_j)(\sigma_z\otimes\sigma_0)\rho(\sigma_z\otimes\sigma_0)(\sigma_i\otimes\sigma_j)$ for any $i,j$. We always take phase shifts to be applied before noise.}.

In addition to spin decoherence, we must account for coupling and transmission losses, inefficient photodetectors, and imperfections in the source. Transmission loss leads to a survival probability of each photon of $\eta_t^S = e^{-L/2L_{att}}$ (symmetric) or $\eta_t^A = e^{-L/L_{att}}$ (asymmetric protocol), where $L_{att}$ is the attenuation length of the channel. The heralding detectors have efficiency $\eta_{her}$. The protocol is also affected by imperfections in the Bell test measurements. The spin readout efficiency can be very high, and so this is not a problem in the symmetric protocol, but in the asymmetric protocol, the efficiency $\eta_d$ of Alice's optical measurement must be considered. Coupling inefficiencies can be absorbed in $\eta_{her}$ and $\eta_d$. For the source, experimentally accessible techniques, such as spontaneous parametric down-conversion, do not generate ideal photonic Bell states, but rather states of the form
\begin{equation}
\label{eq.instate}
\vack + \sqrt{p} \ket{\Phi^+} + O(p) ,
\end{equation}
where $p$ is the probability of generating a photon pair. To avoid errors introduced by multi-photon contributions, $p$ must be kept small, which affects the rate of key generation. In our calculations, we include the leading order of multiphoton terms and optimise $p$ to maximise the key rate.

\begin{figure}[t]
\includegraphics[width = 0.95\linewidth]{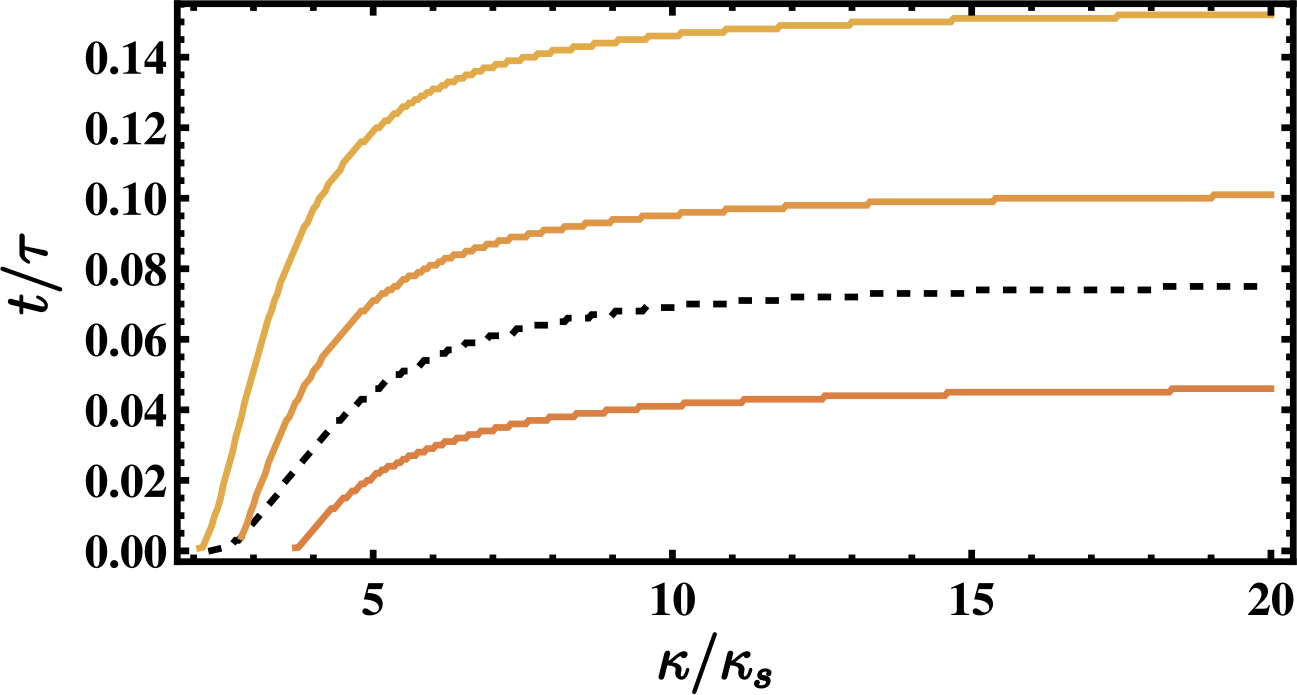}
\caption{A positive key rate can be obtained in the region under the curves for the symmetric (dotted) and the asymmetric protocol with $\eta_d = 1.0$, 0.9, and 0.8 (solid, top to bottom). Note that beyond $\kappa/\kappa_s \sim 10$ increasing $\kappa/\kappa_s$ does not improve the range for $t/\tau$ significantly.}\label{fig.paramranges}
\end{figure}

The figure of merit for a QKD protocol is the key rate, i.e.~how many bits of secret key can be generated per unit time. In the following, for the sake of comparison we will apply the same security analysis as in Ref.~\cite{Gisin2010}, which is based on the Clauser-Horne-Shimony-Holt (CHSH)Bell inequality and which provides security against collective attacks of an eavesdropper. We note however that more recent methods provide security against more general attacks, and enable the computation of key rates for protocols based on other Bell inequalities \cite{masanes2011}. In the future, we will provide the key rate of our scheme also under general attacks. From the observed values of the quantum bit-error rate, we expect corrections by less than an order of magnitude.

The key rate under collective attacks can be computed as a product of the repetition rate $r$ of the source, the total probability for successful heralding and Bell test measurements, and a factor which depends on the CHSH violation and observed correlations. Following \cite{Pironio2009} and the supplementary material of \cite{Gisin2010}, for a given state, the last factor of the key rate is
\begin{equation}
R(\mu,Q,S) = 1-h(Q)-\left[(1-\mu)\chi\left(\frac{S-4\mu}{1-\mu}\right)+\mu\right] ,
\end{equation}
where $h$ is the binary entropy, $\chi(S)=h\left((1+\sqrt{(S/2)^2-1})/2\right)$, and $\mu$ is the ratio of measurement events with indefinite (no click) to those with definite (click) outcomes, i.e.~$\mu^S=0$ while $\mu^A \approx (1-\eta_d)/\eta_d$ for small $p$ (with weak dependence on $L/L_{att}$, $\eta_{her}$, $\kappa/\kappa_s$, and $p$). $S$ and $Q$ are respectively the maximal CHSH violation and minimum  quantum bit-error rate (the probability that Alice's and Bob's outcome are not correlated). $S$ can be computed using the method of the Horodecki's \cite{Horodecki2009} and $Q$ in a similar manner. Intuitively, $R$ is the difference between Alice's and Bob's mutual information and the maximal information that an eavesdropper Eve can extract at each round. Thus, as long as $R$ is positive, Eve cannot extract complete information about the data shared by Alice and Bob, and secrecy is guaranteed. \figref{fig.paramranges} shows the regions of parameters that allow a positive key rate. For sufficiently large detection efficiency $\eta_d$, the asymmetric protocol tolerates more spin decoherence, i.e.~larger $t/\tau$, which can be understood intuitively since only one spin, rather than two, decoheres.

\begin{figure}[t]
\includegraphics[width = .95\linewidth]{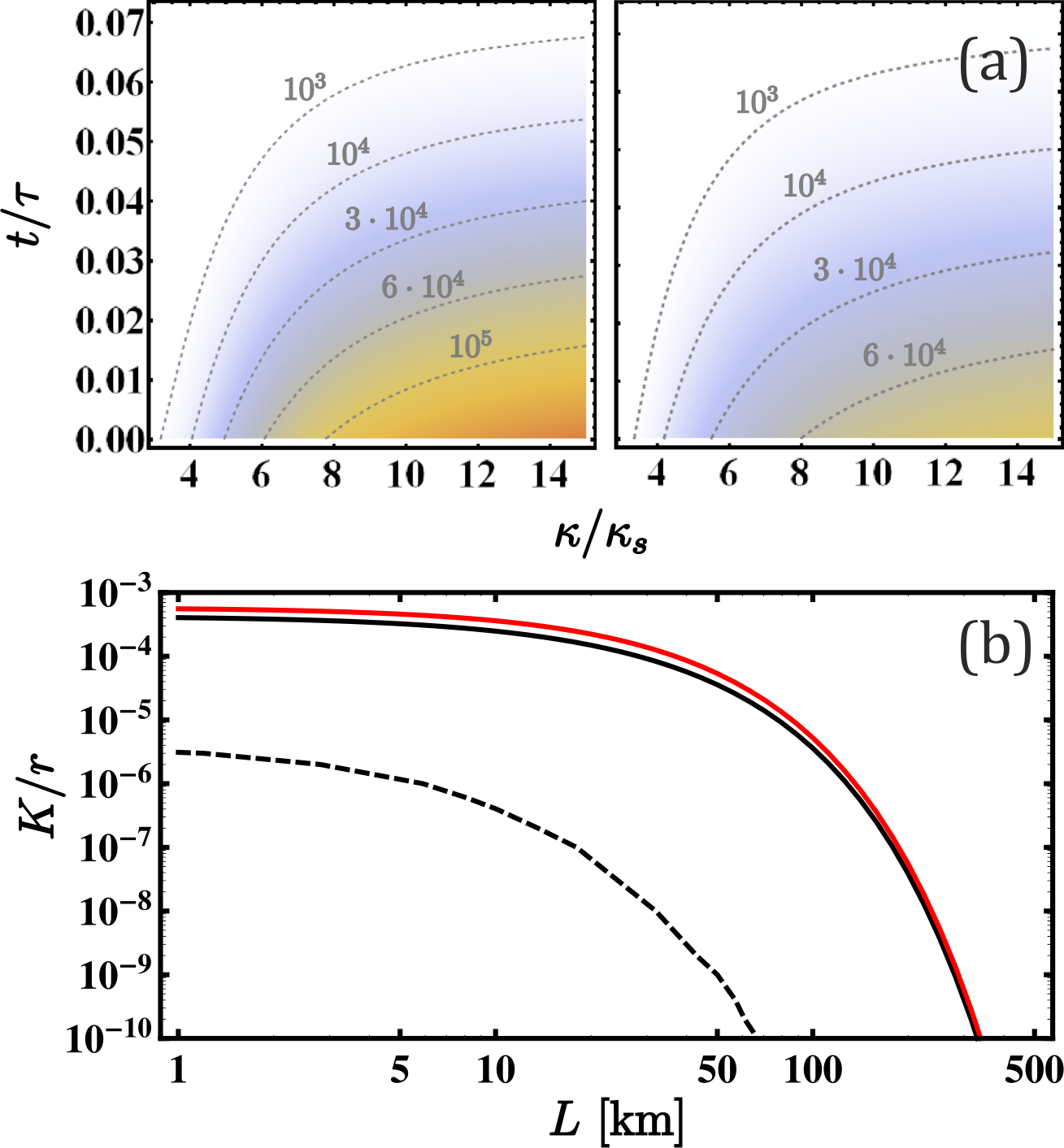}
\caption{\textbf{(a)} Rates of the symmetric (left) and asymmetric (right) protocols at $L=10$km with $r=100$Mhz (contours in bits/s). \textbf{(b)} Key rate in units of the repetition rate for the symmetric (red) and asymmetric (black) protocol with $\kappa/\kappa_s = 6$, $t/\tau = 0.01$. For reference, the rate of Ref.~\cite{Gisin2010} with joint coupling and detection efficiency $0.855$ is also shown (black dashed). For both plots $\eta_{her}=\eta_d=0.855$ and $L_{att}=22$km.}
\label{fig.keyrate}
\end{figure}

What parameter values are realistic? In \cite{Young2013} a detailed analysis, inspired by the quantum dot pillar microcavity experiment of \cite{Reithmaier2004}, showed that lowering the number of DBR mirror pairs relative to a strong coupling regime, resonance scattering is achievable with $g=80 \mu eV$, $\gamma=10 \mu eV$, $\kappa_s= 180 \mu eV$ and $\kappa = 2.38 meV$, which yields $\kappa/\kappa_s \approx 13$. Ref.~\cite{Brunner2013} estimates values of $\kappa/\kappa_s \approx 6$, $3$, and $0.43$ for strongly coupled atoms \cite{Ritter2012}, NV-centers \cite{park2006}, and quantum dots \cite{young2011} respectively, and $2$ for NV-centers in low-Q photonic crystal cavities \cite{RiedrichMoller2012}. For the spin coherence vs.~readout time, Ref.~\cite{Brunner2013} estimates that $t/\tau$ could go as low as $10^{-4}$ for atoms and NV-centers, $10^{-3}$ for low-Q cavities and $10^{-1}$ for quantum dots. Entangled source repetition rates may go as high as $10$GHz \cite{zhang2007,Gisin2010}, however for interaction with dipoles in cavities, the source bandwidth is limited by the narrow cavity linewidth, and the rate is reduced. Ref.~\cite{Brunner2013} estimates source repetition rates of $0.1$MHz for atoms and NV-centers and many MHz for quantum dots.

\figref{fig.keyrate} shows the key rates of the symmetric and asymmetric protocols for selected parameter values. For the sake of comparison with the optical DIQKD-protocol in Ref.~\cite{Gisin2010}, we take  $\eta_{her}=\eta_d=0.9\cdot0.95$ as this corresponds to the joint coupling and detection efficiency used there. \figref{fig.keyrate}(a) shows the performance for varying ratios of measurement to decoherence time and coupling to loss, assuming a repetition rate of 100Mhz. The symmetric protocol performs better than the asymmetric one in this case. However, for larger $\eta_d$, as it is apparent from \figref{fig.paramranges} that the asymmetric protocol tolerates larger $t/\tau$, it must outperform the symmetric for sufficiently large decoherence. In \figref{fig.keyrate}(b) we take $\kappa/\kappa_s=6$ and $t/\tau=0.01$ and plot the keyrate measured in units of the repetition rate for our protocols as well as \cite{Gisin2010}. The present protocol delivers a significant improvement in terms of key bits per use of the source.  E.g.~at 75km we gain about 5 orders of magnitude. As mentioned, the repetition rates in the present scheme is more limited than for the purely optical scheme, because of the need to match the cavity linewidth. However, taking e.g.~ $r=10$GHz for \cite{Gisin2010} and $r=100$MHz for the present scheme one still finds a key rate of about $10^3$bits/s here, whereas the optical scheme achieves less than 1bit/s.

In summary, we have presented a scheme for device-independent quantum key distribution based on interaction between light and spins in cavities. We have analysed the achievable key rates taking into account spin decoherence as well as optical losses, and we have shown that current state-of-the-art systems reach promising numbers, making the system considered here a good candidate for experimental implementation of DI-QKD. We remark that a scheme for heralded mapping of photonic entanglement onto atoms in free space has recently been proposed for a Bell test and can be readily adapted to DI-QKD along the same lines as the scheme presented here \cite{Sangouard2013}. It would be interesting to compare the key rates achievable with that scheme to the present ones.

\textit{Note added:} while finishing the present manuscript it has come to our attention that a new all-optical proposal for DIQKD has very recently appeared \cite{Meyer-Scott2013}.

This work was supported by the European EU PERCENT ERC Starting Grant, the Spanish FIS2010-14830 and Chist-Era DIQIP projects, CatalunyaCaixa, and the Mexican CONACyT Graduate Fellowship Program.

\bibliography{diqkd_dots}

\end{document}